\documentclass[epj,twocolumn]{webofc}
\usepackage[varg]{txfonts}   
\usepackage{hyperref}
\woctitle{Flavour changing and conserving processes}
\begin{document}
\title{On the Hadronic light-by-light contribution to the muon $g-2$}
%
%

\author{Johan Bijnens\inst{1}\fnsep\thanks{\email{bijnens@thep.lu.se}}
}

\institute{Department of Astronomy and Theoretical Physics, Lund University, S\"olvegatan 14A, SE22362 Lund, Sweden
          }

\abstract{
This talk is about the hadronic light-by-light contribution
to the muon anomalous magnetic moment, mainly our old work but including
some newer results as well. It concentrates on the model calculations.
Most attention is paid to pseudo-scalar exchange and the pion loop
contribution. Scalar, $a_1$-exchange and other contributions are shortly
discussed as well.
For the $\pi^0$-exchange a possible large cancellation between connected
and disconnected diagrams is expected.
} 
%
\onecolumn
\thispagestyle{empty}
\begin{flushright}
\large
LU TP 17-43\\
December 2017
\end{flushright}

\vfill

\begin{center}
{\huge\bf
On the Hadronic light-by-light contribution to the muon $g-2$}\\[1.5cm]
{\large\bf Johan Bijnens}\\[1cm]
{\large Dept. of Astronomy and Theoretical Physics, Lund University,\\[1mm]
S\"olvegatan 14A, 22362 Lund, Sweden}

\vfill

{\large\bf Abstract}\\[3mm]

\begin{minipage}{0.8\textwidth}
This talk is about the hadronic light-by-light contribution
to the muon anomalous magnetic moment, mainly our old work but including
some newer results as well. It concentrates on the model calculations.
Most attention is paid to pseudo-scalar exchange and the pion loop
contribution. Scalar, $a_1$-exchange and other contributions are shortly
discussed as well.
For the $\pi^0$-exchange a possible large cancellation between connected
and disconnected diagrams is expected.
\end{minipage}
\end{center}
\vfill
\noindent\rule{8cm}{0.5pt}\\
$^*$ Invited talk  FCCP2017 -
 Workshop on ``Flavour changing and conserving processes,'' 7-9 September
2017,
Anacapri, Capri Island, Italy.
\setcounter{page}{0}
\newpage
\twocolumn
\maketitle
\section{Introduction}
\label{intro}
\begin{figure}
\sidecaption
\includegraphics[width=4.5cm,clip]{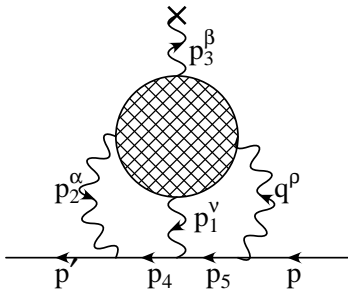}
\caption{HLbL contribution to the muon anomaly $a_\mu$. The crossed circle
indicates the strong interaction part.}
\label{figHLBL}
\end{figure}

This talk is mainly an update of my talk from two years ago \cite{FCCP15}
and has thus a very large overlap with it.
In addition, this writeup should be read together with a number of other
contributions to this and the previous workshop \cite{Proceedings:2016bng}.
A more general introduction to the muon anomaly $a_\mu=(g_\mu-2)/2$ was
given in the talk by and Knecht \cite{Knecht}.
An alternative method to obtain the hadronic light-by-light-contribution (HLbL)
shown in Fig.~\ref{figHLBL} was discussed in the talk by
Colangelo \cite{Colangelo}. The present status of the lattice calculations
of the same quantity were discussed by Lehner \cite{Lehner} and
Nyffeler \cite{Nyffeler}.

The main reason for this sessions is the measurement of the muon anomalous
magnetic moment of \cite{Bennett:2006fi} and the discrepancy with the
standard model prediction and the new measurements in progress at Fermilab
and under development at J-PARC.
Reviews of the theory can be found in
\cite{Bijnens:2007pz,Prades:2009tw,Jegerlehner:2009ry}.
More references can be found in the remainder of this talk and the
talks mentioned above. The present best estimate
of the HLbL is $(11\pm4)\times10^{-10}$ \cite{Bijnens:2007pz,Jegerlehner:2009ry}
or $(10.5\pm2.6)\times10^{-10}$ \cite{Prades:2009tw}.
The main difference is an estimate of the errors which is always somewhat
subjective. A new report with the aim of getting a consensus is under
way as discussed in \cite{Lehner2}.

In this talk I will concentrate on the work done a long time ago
\cite{Bijnens:1995cc,Bijnens:1995xf,Bijnens:2001cq} as well
as some newer work on the pion loop \cite{Bijnens:2016hgx}. I will also discuss
more recent contributions about the pseudo-scalar exchange and
quark-loop. I do not present a new final overall number but will
argue that a good estimate for the pion-loop contribution is
$-(2.0\pm0.5)\times10^{-10}$.

An often asked question is why one cannot simply calculate the
hadronic parts in Chiral Perturbation Theory (ChPT). ChPT is an effective
field theory approximation to QCD valid at low energies. Since the muon has a
low mass, at first sight $a_\mu$ should be a perfect quantity to calculate in
ChPT. This is not true while both for HLbL and the hadronic vacuum
polarization contribution (HVP) there are integrals over all photon momenta
present. The hadronic part is thus not only at low-energies. The lowest-order
prediction for both HVP and HLbL is the same as for scalar QED and is finite.
However, higher orders require a higher dimensional counterterm that is
precisely the same as the muon Pauli term. We are thus left without a
prediction beyond lowest order in ChPT. However, ChPT can (and should be)
used to put as many constraints as possible on the underlying hadronic
quantities.

We thus need to go beyond ChPT since we need high energies and beyond
perturbative QCD since we need low energies for the hadronic quantities.
The main options are experiment, dispersion relations, lattice QCD and models.
For the HVP contributions models only play a role in understanding the
results from the other approaches. For HLbL we have not quite reached that
stage but important progress is being made as discussed by Colangelo, Lehner
and Nyffeler. In the future, the main roles for models will be estimating
the contributions that are not included in the systematic approaches.

The requirement for a model calculation is simple to formulate
``do as well as you can.'' That means constraining your model as much as
possible from experiment via measured states, form-factors and scattering
processes and from theory by including as many long-distance constraints from
ChPT and short-distance constraints from perturbative QCD. One should also use
``common sense'' in varying model parameters, making sure your model is general
enough to describe what you need to describe and if different regions
are treated differently consistency between them should be checked.

An overview of general properties of the underlying four-point functions
and the early calculations is in Sect.~\ref{general}.
Sect.~\ref{pi0exchange} discusses the numerically largest
contribution, pseudo-scalar meson exchange. 
Next I discuss the pion-loop contribution in some detail since here I have
new results \cite{Bijnens:2016hgx}.
The quark-loop, which has rather large theoretical errors, is discussed in
Sect.~\ref{quarkloop}.
The remaining leading large $N_c$ exchanges are
scalar, discussed in Sect.~\ref{scalar}, and $a_1$-exchange, Sect.~\ref{a1}.
The $\pi$-loop contribution is treated in more detail
since there is where I have some new results to present. Details are
in Sect.~\ref{piloop}. Conclusions and some possible future
directions are given in the last section.

\section{General properties and early work}
\label{general}

The underlying object is the four-point function
$\Pi^{\rho\nu\alpha\beta} (p_1,p_2,p_3)$ of four electromagnetic vector currents.
We really need only the derivative w.r.t. $p_3$ at $p_3=0$,
\begin{equation}
 \left. {\delta \Pi^{\rho\nu\alpha\beta}(p_1,p_2,p_3)\over \delta  p_{3\lambda}}\right|_{p_3=0}\,.
\end{equation}
$\Pi^{\rho\nu\alpha\beta} (p_1,p_2,p_3)$ has in general 138 Lorentz structures
which reduces to 43 gauge-invariant structures. Note that
in four dimensions there really are 2 less, 136 and 41 \cite{Eichmann:2014ooa}.
Of the 138 more general structures 28 \cite{Bijnens:2016hgx} actually
contribute (improving the 32 estimate of \cite{Bijnens:1995xf}).
Each of these functions depends on $p_1^2,p_2^2,q^2$
and before the derivative also on $p_3^2,p_1.p_3,p_2.p_3$.
This should be compared with the lowest order hadronic vacuum polarization
where there is one function of one variable.
An alternative split is using the helicity amplitudes for off-shell
photon-photon scattering as used in the dispersive work by Colangelo and
collaborators \cite{Colangelo}. The choice of basis is definitely not unique
and different choices are appropriate for the different approaches.

After setting $p_3\to0$ the loop integrals over the photon momenta
is 8 dimensional. Three of these integrations are trivial and using
Gegenbauer polynomial methods two more can be done
\cite{Jegerlehner:2009ry,Bijnens:2016hgx,Knecht:2001qf}. So, after having a model or a
computation of $\Pi^{\rho\nu\alpha\beta} (p_1,p_2,p_3)$ there is a triple integral
over $p_1^2,p_2^2,q^2$ left. The components and their derivatives
become multiplied with functions of $p_1^2,p_2^2,q^2$ examples of which
are in \cite{Jegerlehner:2009ry,Knecht:2001qf} and the full results can be
found in \cite{Bijnens:2016hgx}. In the work I have been involved in we did
the relevant integrations in Euclidean space, i.e. with $P_1^2,P_2^2,Q^2=
-p_1^2,-p_2^2,-q^2$ always positive.

How models actually contribute to the muon anomaly $a_\mu$
can be studied by rewriting the integral over $P_1^,P_2^2,Q^2$ in the form
\cite{Bijnens:2007pz}
\begin{equation}
\label{defaLL}
a_\mu = \int dl_{P_1} dl_{P_2}\, { a_\mu^\mathrm{LL}}
 = \int dl_{P_1} dl_{P_2} dl_Q\, { a_\mu^\mathrm{LLQ}}
\end{equation}
with $l_P = (1/2)\ln\left(P^2/\mathrm GeV^2\right)$. The reason for choosing the
logarithm is that this way it is easiest to see which momentum region
contributes. Alternatively one can integrate each momentum up to a
cut-off $\Lambda$.

One should remember that the different contributions are usually defined within
a given model or approach. What is included under the same name can therefore
differ and one should be careful when drawing conclusions from comparing
calculations.

The underlying problem is that the integration over photon momenta
$p_1,p_2$ in the diagram in Fig.~\ref{figHLBL} contains both low and high
momenta and mixed cases. Double counting is thus a serious issue when
using both quark and hadron contributions. In Ref.~\cite{deRafael:1993za}
a partial solution was found by using chiral $p$ and large $N_c$ counting
to distinguish different contributions. This does not fully solve the
double counting issue but it is a good start. This suggestion was followed by
two groups doing a more or less full evaluation of the HLbL.
Kinoshita and collaborators
\cite{Hayakawa:1995ps,Hayakawa:1996ki,Hayakawa:1997rq} (HKS) used meson models,
did the pion-loop using the hidden local symmetry model for vector mesons
and the quark loop with simple vector meson dominance (VMD).
Calculations were performed in Minkowski space.
The one I was
involved in \cite{Bijnens:1995cc,Bijnens:1995xf,Bijnens:2001cq} (BPP)
tried to use a consistent model, the extended Nambu-Jona-Lasinio (ENJL) model
as in \cite{Bijnens:1992uz,Bijnens:1995ww},
as much as possible but adjusted using measured form-factors and QCD
constraints. The calculations were done in Euclidean space.
In fact, these two are still the only existing full calculations, but many parts
have been evaluated using other approaches since then.

The main observations were:
\begin{itemize}
\item The largest contribution is $\pi^0$ (and $\eta,\eta^\prime$)
 exchange/pole. Be aware that exchange and pole or not precisely the same.
Most estimates of this part are in reasonable agreement as discussed in
Sect.~\ref{pi0exchange}.
\item The pion loop can be sizable, with a large difference between the
two evaluations and even larger numbers have been proposed. Further discussion
is in Sect.~\ref{piloop}.
\item The other contributions are smaller but there are many and cancellations
are present.
\item Final numbers:\\BPP: $(8.3\pm3.1)~10^{-10}$, HKS: $(8.96\pm1.54)~10^{-10}$.
\end{itemize}

\section{\boldmath$\pi^0$-exchange}
\label{pi0exchange}

\begin{figure}
\sidecaption
\includegraphics[width=4cm]{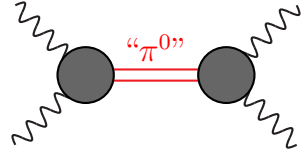}
\caption{The $\pi^0$ exchange contribution. The blobs and the
propagator need modeling.}
\label{figpi0}
\end{figure}

The single largest numerical contribution is given by ``$\pi^0$'' exchange,
depicted in Fig.~\ref{figpi0}.
The blobs need modeling and the propagator in the ENJL model also has
corrections to the $1/(p^2-m_{\pi^0}^2)$. The pointlike vertex has a logarithmic
divergence which is uniquely predicted \cite{Knecht:2001qg,RamseyMusolf:2002cy}.
The VMD form-factor in the $\pi^0\gamma^*\gamma^*$ form-factor, the blobs,
were modeled in \cite{Bijnens:1995xf} with a variety of form-factors and
as a function of the cut-off $\Lambda$ (corrected for the overall sign error
discovered by \cite{Knecht:2001qf}).
\begin{table}
\caption{The $\pi^0$ exchange results of \cite{Bijnens:1995xf}.} 
\label{tabpi0}
\centering
\begin{tabular}{|c|c@{ }c@{ }c@{ }c@{ }c|}
\hline &\multicolumn{5}{c|}{$a_\mu\times 10^{10}$}
\\
\hline
$\Lambda$ & Point- & ENJL-&Point- &Transv. &CELLO-\\
GeV       & like   & VMD  & VMD   & VMD    & VMD\\
\hline
0.5  & 4.92(2)& 3.29(2)  & 3.46(2) &3.60(3)&3.53(2)\\
0.7  &7.68(4)& 4.24(4)  & 4.49(3) &4.73(4)&4.57(4)\\
1.0  & 11.15(7)&4.90(5)  & 5.18(3)&5.61(6)&5.29(5)\\
2.0  & 21.3(2)& 5.63(8)  & 5.62(5)&6.39(9)&5.89(8)\\
4.0  &32.7(5)& 6.22(17) & 5.58(5)&6.59(16)&6.02(10)\\
\hline
\end{tabular}
\end{table}
We took the form-factor that was made to fit the then existing data integrated
up to 2~GeV as our main result with a guesstimate of the error.
This result was in quite good agreement with \cite{Hayakawa:1996ki} which used
the pointlike-VMD approach. This contribution has since been reevaluated
many times using different models and approaches. A partial list is:\\
BPP \cite{Bijnens:1995xf}: \hfill $ 5.9(0.9)\times 10^{-10}$\\
Nonlocal quark model \cite{Dorokhov:2008pw}:  \hfill  $ 6.27\times 10^{-10}$\\
DSE (Dyson-Schwinger modeling)\cite{Goecke:2010if}:\hfill
  $5.75\times 10^{-10}$\\
LMD+V \cite{Knecht:2001qf}: \hfill $ (5.8-6.3)\times 10^{-10}$\\
Formfactor inspired by AdS/QCD \cite{Cappiello:2010uy}:\hfill 
$6.54\times 10^{-10}$\\
Chiral Quark Model \cite{Greynat:2012ww}:\hfill $6.8\times 10^{-10}$\\
Constraint via magnetic susceptibility \cite{Nyffeler:2009tw}:\hfill
 $ 7.2\times 10^{-10}$\\
$VV^\prime P$ model \cite{Roig:2014uja}:\hfill $6.66\times 10^{-10}$\\
All of these are in reasonable agreement, within the errors.
Future improvements will come when more experimental results or the direct
lattice calculations of the underlying formfactors are included.
One way to define precisely the $\pi^0$ pole contribution is the dispersive
approach \cite{Colangelo:2014dfa} and the talk \cite{Colangelo}, but no
numerical results have been published so far.

Two more comments are needed. The above numbers are for the $\pi^0$.
One needs to take into account the $\eta$ and $\eta^\prime$ exchange as well.
The latter is enhanced due to the charge combinations in the
$\eta^\prime\gamma^*\gamma^*$ vertex. In large $N_c$ models like the ENJL model,
the pseudoscalar spectrum is not like QCD, one has a $\pi^0$,
a $\tilde\pi$ ($\bar u u+\bar dd$ quark content) and a $\pi_s$ ($\bar ss$).
The $\tilde\pi$ has the same mass as the $\pi^0$ and due to the
quark charges is contributes $25/9$ times the $\pi^0$ contribution. Lattice QCD
calculations with only connected diagrams included will have the $\tilde\pi$
contribution as well so there will be an unphysical enhancement compared
to the QCD result for the pseudoscalar exchange part.
This is discussed in more detail in \cite{Bijnens:2016hgx}.
In \cite{Bijnens:1995xf} we used pointlike-VMD to estimate the ratio
of $\pi^0,\eta,\eta^\prime$ contributions as $5.58,1.38,1.04$. Models
that include large $N_c$-breaking effects and fit the mixings to data typically
end up with very similar numbers. The total pseudoscalar exchange contribution
I thus estimate to be
\begin{equation}
a_\mu^{PS} = (8-10)\times10^{-10}
\end{equation}
An example of a specific calculation is the AdS/QCD result
of  $a_\mu^{PS} = 10.7\times10^{-10}$ \cite{Hong:2009zw} which also includes
excited pseudoscalars.

The other comment is that the short-distance behaviour of the four-point
function is known in several limits. In particular when $P_1^2\approx P_2^2
\gg Q^2$ the four point function is related to the axial-vector-vector-vector
three-point function \cite{Melnikov:2003xd}. This three point function
has a number of exact properties in QCD and we thus know how it behaves.
The above models for $\pi^0$-exchange do not exhibit this behaviour.
It can be implemented via making one of the blobs in
Fig.~\ref{figpi0} pointlike \cite{Melnikov:2003xd} and one then obtains
$7.7\times 10^{-10}$ for the $\pi^0$-exchange contribution. 
Plots how this affects the contribution of different momentum regions are in
\cite{Bijnens:2007pz}.
The above behaviour
of the four-point function must be obeyed in a full calculation, however
whether one implements it via $\pi^0$-exchange is a choice.
Models incorporating a short-distance quark-loop contribution have the
short-distance part of this included \cite{Bijnens:2007pz,Dorokhov:2008pw}.
One can see this when comparing quark-loop plus pseudo-scalar exchange
of \cite{Bijnens:1995xf} with pseudo-scalar exchange of \cite{Melnikov:2003xd}.

\section{\boldmath$\pi$-loop}
\label{piloop}

\begin{figure}
\centering
\includegraphics[width=6cm]{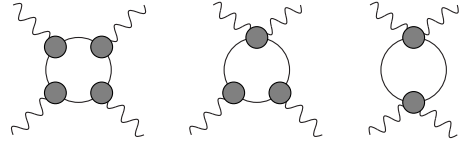}
\caption{The charged pion loop contribution.}
\label{figpiloop}
\end{figure}

The $\pi$-loop contribution to the four-point function is depicted
in Fig.~\ref{figpiloop}.
The leftmost diagram is the naive one, the other two are required by
gauge-invariance. In more general models also a diagram with three photons
in one vertex and one with all four in the same vertex might be needed.
These have been included in the calculations mentioned below when needed.

The simplest model is a point-like pion or scalar QED (sQED). This gives
a contribution of about $-4\times10^{-10}$. 

\begin{figure}
\centering
\includegraphics[width=0.99\columnwidth]{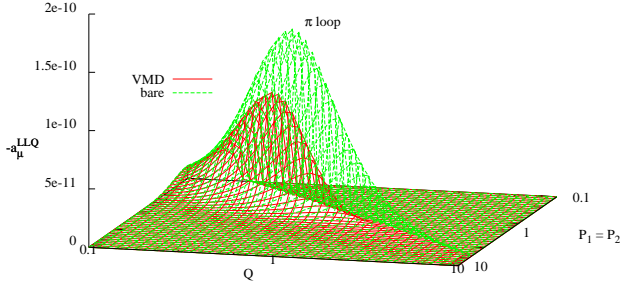}
\caption{The momentum dependence of the pion loop contribution.
Plotted is $a_\mu^{LLQ}$
of (\ref{defaLL})
as a function of $P_1=P_2$ and $Q$. Top surface: sQED, bottom surface:full VMD.}
\label{figpiloopVMD}
\end{figure}

The single photon vertex is in
all determinations used as including the pion form-factor. For this one
can use either the VMD expression or a more model/experimental inspired
version. For the $\pi\pi\gamma^*\gamma^*$ vertex there
were originally two main approaches used,
full VMD (BPP) and the hidden local symmetry model with vector mesons (HKS).
The former is essentially using sQED and putting a VMD-like form-factor
in all the photon legs. This was proven to be a consistent procedure
in \cite{Bijnens:1995xf}. We obtained there a result of $-1.9\times 10^{-10}$
using an ENJL inspired pion form-factor. Using a simple VMD typically gives
about $-1.6\times 10^{-10}$. This version is exactly what is called the
model-independent part of the two-pion contribution in
\cite{Colangelo,Colangelo:2014dfa,Colangelo:2014pva}.
The reason for the lower number compared to the point-like pion loop
is obvious in Fig.~\ref{figpiloopVMD} where we show $a_\mu^{LLQ}$
of (\ref{defaLL})
as a function of $P_1=P_2$ and $Q$.

\begin{figure}
\centering
\includegraphics[width=0.99\columnwidth]{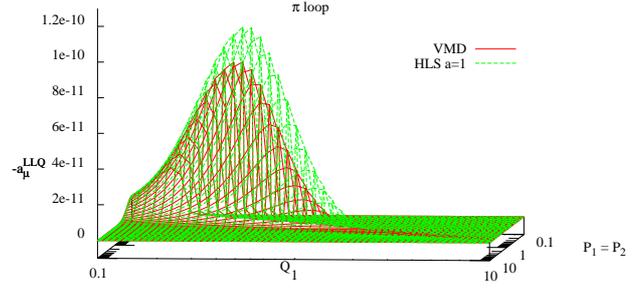}
\caption{The momentum dependence of the pion loop contribution.
$-a_\mu^{LLQ}$
of (\ref{defaLL})
as a function of $P_1=P_2$ and $Q$. Top surface: HLS a=1, bottom surface: full VMD.}
\label{figpiloopVMDHLS2}
\end{figure}

HKS \cite{Hayakawa:1995ps,Hayakawa:1996ki} used a different approach.
Due to the then existing arguments against full VMD they used the hidden local
symmetry model with only vector mesons (HLS) and obtained
$-0.45\times 10^{-10}$. The difference between this and the previous numbers
was the reason for the large error quoted on the pion-loop.
This difference was rather puzzling, one reason could be that the
HLS model does not have the correct QCD short distance constraint when
looking at the two-photon vertex with the same and large virtuality for both
photons, the full VMD model has the correct behaviour.
This version of the HLS model also does not give a finite prediction for
the $\pi^+$-$\pi^0$ mass difference.
The reason for the large numerical difference is indeed the short distance
behaviour. The low momentum behaviour is very close but the negative
\begin{figure}
\centering
\includegraphics[width=0.99\columnwidth]{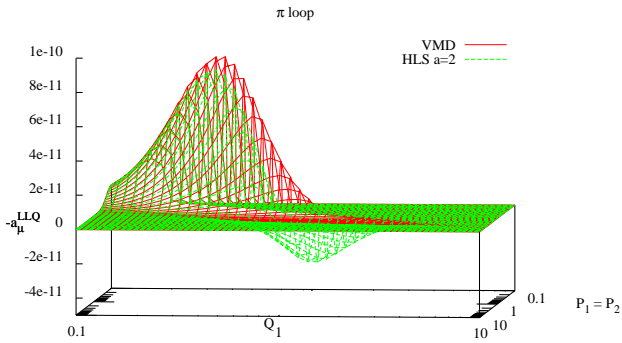}
\caption{$-a_\mu^{LLQ}$
of (\ref{defaLL})
as a function of $P_1=P_2$ and $Q$. Top surface: full VMD, bottom surface: HLS.}
\label{figpiloopVMDHLS}
\end{figure}
contribution above 1~GeV, clearly visible in Fig.~\ref{figpiloopVMDHLS},
is the main reason for the difference
\cite{Bijnens:2012an,Bijnens:2016hgx}. A comparison as a function of the
cut-off can be found in \cite{Abyaneh:2012ak}. In fact, using the HLS with an
unphysical value of the parameter $a=1$, which then satisfies the
abovementioned short-distance constraint gives very similar numbers as full VMD.
This is shown in Fig.~\ref{figpiloopVMDHLS2}
From this we conclude that a number in the range $-(1.5$-$1.9)\times10^{-10}$
is more appropriate with an error of half to 1/3 that.

\begin{figure}
\centering
\includegraphics[width=0.99\columnwidth]{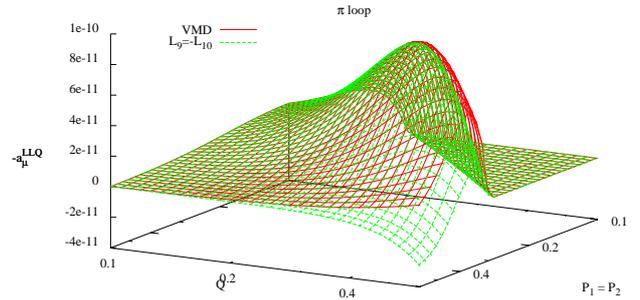}
\caption{$-a_\mu^{LLQ}$
of (\ref{defaLL})
as a function of $P_1=P_2$ and $Q$. Top surface: full VMD, bottom surface: 
ChPT with $L_9=-L_{10}$ so the charge radius is included but no polarizability.}
\label{figpiloopChPT1}
\end{figure}

\begin{figure}
\centering
\includegraphics[width=0.99\columnwidth]{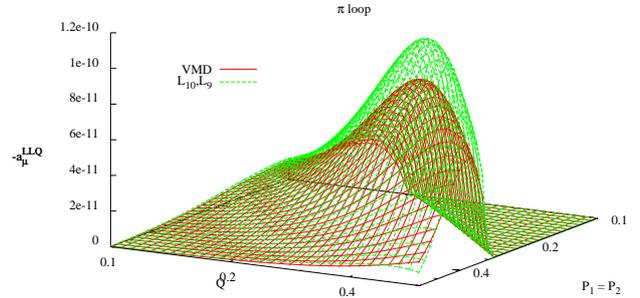}
\caption{$-a_\mu^{LLQ}$
of (\ref{defaLL})
as a function of $P_1=P_2$ and $Q$. Bottom surface: full VMD, top surface: 
ChPT with $L_9\ne-L_{10}$ so the charge radius and the polarizability
are included.}
\label{figpiloopChPT2}
\end{figure}

More recently, it was pointed out that the effect of pion polarizability
was neglected in these calculations and a first estimate of this
effect given
using the Euler-Heisenberg four photon effective vertex produced
by pions \cite{Engel:2012xb} within Chiral Perturbation Theory.
This approximation is only valid
below the pion mass. In order to check the size of the pion radius effect
and the polarizability we have implemented the low energy part
of the four-point function and computed $a_\mu^{LLQ}$ for these cases.
Partial results are in \cite{Bijnens:2012an,Abyaneh:2012ak} and
the full results in \cite{Bijnens:2016hgx}.
The effect of the charge radius is shown in Fig.~\ref{figpiloopChPT1}
compared to the VMD, notice the different momentum scales compared to the
earlier figures. As expected, the charge radius effect is included in the
VMD result since the latter gives a good description of the pion form-factor.
Including the effect of the polarizability can be done in ChPT by
using experimentally determined values for $L_9$ and $L_{10}$. The latter
can be determined from $\pi^+\to e\nu\gamma$ or the hadronic vector
two-point functions. Both are in good agreement and lead to a prediction
of the pion polarizability confirmed by the compass experiment
\cite{Adolph:2014kgj}. The effect of including this in ChPT on $a_\mu^{LLQ}$
is shown in Fig.~\ref{figpiloopChPT2}
\cite{Bijnens:2016hgx,Bijnens:2012an,Abyaneh:2012ak}. An increase of 10-15\% over the
VMD estimate can be seen.

ChPT at lowest order or $p^4$ for $a_\mu$ is just the pointlike pion loop
or sQED. At NLO pion exchange with pointlike vertices and the pionloop
calculated at NLO in ChPT are needed. Both gives divergent contributions
to $a_\mu$, so pure ChPT is of little use in predicting $a_\mu$. 
If we want to see
the full effect of the polarizability we need to include a model that can be
extended all the way, or at least to a cut-off of about 1~GeV.
For the approach of \cite{Engel:2012xb} this was done in 
\cite{Engel:2013kda} by including a propagator description of $a_1$
and choosing it such that the full contribution of the pion-loop
to $a_\mu$ is finite. They obtained a range of $-(1.1$-$7.1)\times 10^{-10}$
for the pion-loop contribution. I find this range much too broad.
One reason is that the range of polarizabilities used in \cite{Engel:2013kda}
is simply not compatible with ChPT. The pion polarizability is an observable
where ChPT should work and indeed the convergence is excellent. The ChPT
prediction has also recently been confirmed by experiment. Our work 
discussed below indicates that $-(2.0\pm0.5)\times10^{-10}$ is a more
appropriate range for the pion-loop contribution.

\begin{figure}
\centering
\includegraphics[width=0.9\columnwidth]{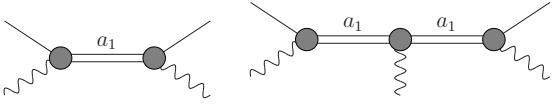}
\caption{Left: the $a_1$-exchange that produces the pion polarizability.
Right: an example of a diagram that is required by gauge invariance.}
\label{figa1pipi}
\end{figure}

The work described below is be published in \cite{Bijnens:2016hgx}.
Preliminary results have been reported at several conferences, see e.g.
\cite{Amaryan:2013eja,Benayoun:2014tra}.
The polarizability comes from $L_9+L_{10}$ in ChPT.
Using \cite{Ecker:1988te}, we notice that the polarizability is produced by
$a_1$-exchange depicted in Fig.~\ref{figa1pipi}. This is depicted pictorially
in the left diagram of Fig.~\ref{figa1pipi}.
However, once such an exchange is there, diagrams like the right one in
Fig.~\ref{figa1pipi} lead to effective $\pi\pi\gamma\gamma\gamma$ vertices
and are required by electromagnetic gauge invariance. This was done
in \cite{Engel:2013kda} via the propagator modifications. We deal with them via
effective Lagrangians incorporating vector and axial-vector mesons.

If one looks at Fig.~\ref{figa1pipi} one could raise the question ``Is
including a $\pi$-loop but no $a_1$-loop consistent?''
The answer is yes with the following argument. We can first look at a tree
level Lagrangian including pions $\rho$ and $a_1$. We then integrate out the
$\rho$ and $a_1$ and calculate the one-loop pion diagrams
with the resulting Lagrangian.
In the diagrams of the original Lagrangian this corresponds to only including
loops with at least one pion propagator present. Numerical results for cases
including full $a_1$ loops are presented as well below \cite{Bijnens:2016hgx}.
As a technicality, we use anti-symmetric vector fields for the vector and
axial-vector mesons. This avoids complications due to $\pi$-$a_1$ mixing.
We add vector $V_{\mu\nu}$ and axial-vector $A_{\mu\nu}$ nonet fields.
The kinetic terms are given by \cite{Ecker:1988te}
\begin{equation}
-\frac{1}{2}\left\langle\nabla^\lambda V_{\lambda\mu}\nabla_\nu V^{\nu\mu}
-\frac{M_V^2}{2}V_{\mu\nu}V^{\mu\nu}\right\rangle
+ V\leftrightarrow A\,.
\end{equation}
First we add the terms that contribute to the $L_i$ \cite{Ecker:1988te}
\begin{equation}
 \frac{F_V}{2\sqrt{2}}\left\langle f_{+\mu\nu}V^{\mu\nu}\right\rangle
+\frac{i G_V}{\sqrt{2}}\left\langle V^{\mu\nu}u_\mu u_\nu\right\rangle
+\frac{F_A}{2\sqrt{2}}\left\langle f_{-\mu\nu}A^{\mu\nu}\right\rangle
\end{equation}
with
$L_9 = \frac{F_V G_V}{2 M_V^2}$, $L_{10}=-\frac{F_V^2}{4M_V^2}+\frac{F_A^2}{4M_A^2}$. The Weinberg sum rules imply in the chiral limit
$F_V^2 = F_A^2 + F^2_\pi$, $F_V^2 M_V^2 = F_A^2 M_A^2$
and requiring VMD behaviour for the pion form-factor $ F_V G_V = F^2_\pi$.

First, look at the model with only $\pi$ and $\rho$.
The one-loop contributions to $\Pi^{\rho\nu\alpha\beta}$
are not finite. They were also not finite for the HLS
model of HKS, but the relevant
 $\delta\Pi^{\rho\nu\alpha\beta}/\delta p_{3\lambda}$ was. However, in the present
model it is only finite for $G_V=F_V/2$ and then the result for $a_\mu$
is identical to the HLS model. The same comments as made for the HLS model
thus also apply.

\begin{figure}
\centering
\includegraphics[width=0.95\columnwidth]{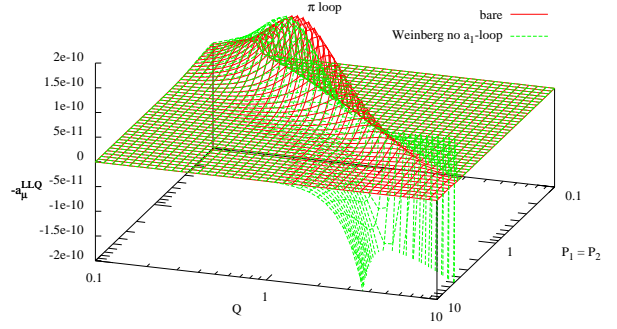}
\caption{$-a_\mu{LLQ}$ as defined in (\ref{defaLL}) as a function of
$P_1=P_2$ and $Q$ with $a_1$ but no full $a_1$-loop. Parameters determined by the Weinberg sum rules.}
\label{figa1LLQ}
\end{figure}

\begin{figure}
\centering
\includegraphics[width=0.95\columnwidth]{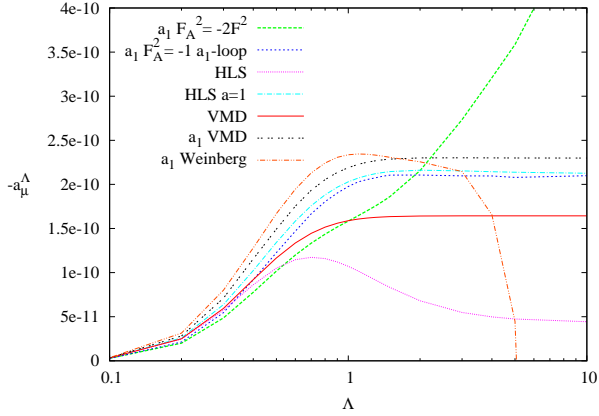}
\caption{$-a_\mu$ using a variety of models for the pion loop as a function of $\Lambda$, the cut-off on the photon momenta. Figure from \cite{Bijnens:2016hgx}.}
\label{figpiloopall}
\end{figure}

Next we do add the $a_1$ and require $F_A\ne0$. After a lot of work we find
that
$\delta\Pi^{\rho\nu\alpha\beta}/\delta p_{3\lambda}|_{p_3=0}$ is finite only for
$G_V=F_V=0$ and $F_A^2=-2F_\pi^2$ or, if including a full $a_1$-loop
$F_A^2=-F_\pi^2$. These solutions are clearly unphysical.
We then add all $\rho a_1\pi$ vertices given by
\begin{align}
&\lambda_1\left\langle \left[V^{\mu\nu},A_{\mu\nu}\right]\chi_-\right\rangle
+\lambda_2\left\langle \left[V^{\mu\nu},A_{\nu\alpha}\right]{h_\mu}^\nu\right\rangle
\nonumber\\&
+\lambda_3\left\langle i\left[\nabla^\mu V_{\mu\nu},A_{\nu\alpha}\right]u_\alpha\right\rangle
+\lambda_4\left\langle i\left[\nabla_\alpha V_{\mu\nu},A_{\alpha\nu}\right]u^\mu\right\rangle
\nonumber\\&
+\lambda_5\left\langle i\left[\nabla^\alpha V_{\mu\nu},A_{\mu\nu}\right]u_\alpha\right\rangle
+\lambda_6\left\langle i\left[V^{\mu\nu},A_{\mu\nu}\right]{{f_{-}}^\alpha}_\nu\right\rangle
\nonumber\\&
+\lambda_7\left\langle i V_{\mu\nu}A^{\mu\rho}{A^\nu}_\rho\right\rangle\,.
\end{align}
These are not all independent due to the constraints on $V_{\mu\nu}$ and
$A_{\mu\nu}$ \cite{Leupold}, there are three relations.
After a lot of work \cite{Bijnens:2016hgx} we found that no solutions with
$\delta\Pi^{\rho\nu\alpha\beta}/\delta p_{3\lambda}|_{p_3=0}$ exists except
those already obtained without $\Lambda_i$ terms.
The same conclusions holds if we look at the combination that shows up in the
integral over $P_1^2,P_2^2,Q^2$. We thus find no reasonable model
that has a finite prediction for $a_\mu$ for the pion-loop including
$a_1$. If we choose the parameters as fixed by the Weinberg sum rules and the
VMD behaviour of the pion-form factor we obtain $-a_\mu^{LLQ}$ as shown in
Fig.~\ref{figa1LLQ}.
Adding a full $a_1$-loop changes the plot only marginally.
As long as we require the correct polarizability and a VMD-like form-factor
behaviour,
the plots look quite similar for all cases below 1~GeV.
The integrated value up to $\Lambda$ for a number of cases is how in Fig.~\ref{figpiloopall}.
We see that all models end up with a value of $a_\mu=-(2.0\pm0.5)\times10^{-10}$
when integrated up-to a cut-off of order 1-2~GeV.
We conclude that that is a reasonable estimate for the pion-loop
contribution. The main missing part is the $\pi$-$\pi$ rescattering.

The dispersive approach has numbers that are compatible with the above
and the inclusion of scalar exchange,
$a_\pi^{\pi-loop} = (-2.4\pm0.1)~10^{-10}$.
See the discussion in \cite{Colangelo,Colangelo:2017qdm,Colangelo:2017fiz}.

\section{Quark-loop}
\label{quarkloop}

\begin{figure}
\centering
\includegraphics[width=0.95\columnwidth]{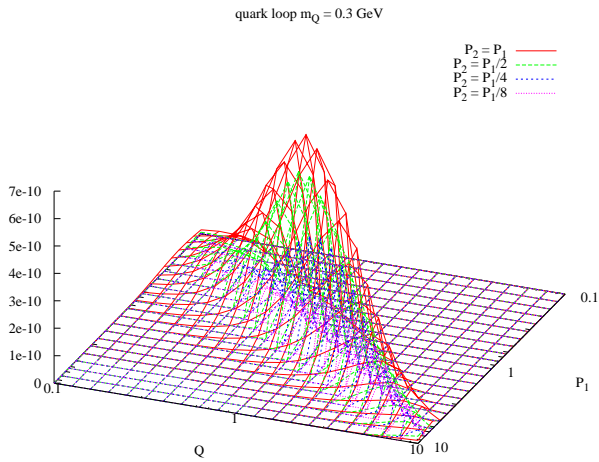}
\caption{The quantity $a_\mu^{LLQ}$ defined in (\ref{defaLL})
as a function of $P_1$ and $Q$ for several ratios $P_1/P_1$. The lines indicate the surface.}
\label{figquarkloop}
\end{figure}

The pure quark-loop contribution with a constant mass is known analytically.
One of the surprises is that it converges rather slowly. A significant portion
is from high momentum regions. With a constituent quark mass of 300~MeV
and a cut-off of 1(2) GeV 50(25)\% of the full contribution is still missing.
A more visual illustration of this is the plot of $a_\mu^{LLQ}$ defined
in (\ref{defaLL}) of this contribution. The contribution is plotted
in Fig.~\ref{figquarkloop}
as a function of $P_1$ and $Q$ for several ratios of $P_2/P_1$.
The volume under the curve is proportional to $a_\mu$. The contribution
peaks for $P_1\approx P_2\approx Q$ and around 1~GeV.
In \cite{Bijnens:1995xf} we used the ENJL up to a cut-off $\Lambda$ and
added a short-distance quark-loop where we used the quark-mass $M_H=\Lambda$
as a lower cut-off. The estimate used by HKS was a quark-loop damped by
VMD factors in the photon legs. The results are given in
Tab.~\ref{tabquarkloop}. Notice especially the stability when we add the
ENJL and the short-distance contribution in the region $\Lambda=1$-8~GeV.
\begin{table}
\caption{The quark-loop contribution with VMD damping, the ENJL model
and with a heavy quark mass as cut-off.
The numbers are $a_\mu\times 10^{10}$.}
\label{tabquarkloop}
\centering
\begin{tabular}{|c|cccc|}
\hline
Cut-off &  &  & & sum \\
$\Lambda$ & &  & mass- & ENJL \\
GeV  & VMD & ENJL & cut & masscut\\
\hline
0.5 & 0.48 & 0.78 & 2.46 & 3.2\\
0.7 & 0.72 & 1.14 & 1.13 & 2.3\\
1.0 & 0.87 & 1.44 & 0.59 &  2.0\\
2.0 & 0.98 & 1.78 & 0.13 &  1.9\\
4.0 & 0.98 & 1.98 & 0.03 &  2.0\\
8.0 & 0.98 & 2.00 & .005 & 2.0\\
\hline
\end{tabular}  
\end{table}
The conclusion is that the quark-loop is about $2\times 10^{-10}$.
In the ENJL model the quark-loop and scalar exchange are needed together
to have correct chiral symmetry. The sum of both is very similar to the
quark-loop estimate of HKS.

There are a number of estimates of the quark-loop that lead to much larger
numbers. These have all in common that there is a momentum region with a fairly
small (constituent) quark mass that is not shielded by a VMD-like mechanism.
The most prominent example of this is the DSE estimate of \cite{Goecke:2012qm}
$10.7(0.2)\times 10^{-10}$. The present status of this calculation
is given in \cite{Eichmann:2014ooa}. It not yet a full calculation but includes
an estimate of some of the missing parts. This DSE model describes a lot of
low-energy phenomenology in a way very similar to the ENJL model. I am quite
puzzled by the difference in results.

Similar size numbers are obtained in models with a low constituent quark
mass where no VMD-like dynamical effects are included.
Examples are the nonlocal chiral quark model \cite{Dorokhov:2015psa}
with $11.0(0.9)\times 10^{-10}$ and a number of
estimates within the chiral quark model
$(7.6-8.9)\times10^{-10}$ \cite{Greynat:2012ww},
$(11.8-14.8)\times10^{-10}$ \cite{Boughezal:2011vw}
and  $(7.6-12.5)\times10^{-10}$\cite{Masjuan:2012qn}. The interpretation
varies from an estimate to the full HLbL or just a part that needs to
be added to other contributions.

\section{Scalar exchange}
\label{scalar}

The estimate of the scalar exchange contribution in the ENJL model
is $-0.7\times10^{-10}$. Similar size estimates have been obtained when
exchanging a sigma-like particle. It should be pointed out that the scalar in
the ENJL model has a phenomenology similar to the sigma but is quite a different
underlying object.

A problem here is to distinguish scalar exchange from two-pion
or pion-loop contributions. This is one of the areas where the
method of \cite{Colangelo,Colangelo:2014dfa} as used
in \cite{Colangelo:2017qdm,Colangelo:2017fiz} allowed for major progress.

\section{\boldmath $a_1$-exchange}
\label{a1}

The exchange of axial vectors in the ENJL model was estimated
in \cite{Bijnens:1995xf} to be about $0.6\times 10^{-10}$,
but due to the high mass involved, even with a cut-off of 2~GeV only half
the contribution was there. The ENJL part also includes some pseudo-scalar meson
exchange due to the structure of the calculation.

Axial-vector meson exchange in a more phenomenological way was done using
two multiplets in \cite{Melnikov:2003xd} who obtained
$2.2\times 10^{-10}$. It was later found that when correct antisymmetrization
is included, this becomes smaller by a significant factor and is again in the
ballpark of the ENJL result.
This was noticed by F.~Jegerlehner. He obtains about
$(0.76\pm0.27)\times 10^{-10}$ for the axial-vector exchange
\cite{Jegerlehner1,Jegerlehner2}.
The evaluation of \cite{Pauk:2014rta} is also in reasonable agreement
with the ENJL estimate.

\section{Conclusions}
\label{conclusions}

The present number for the HLbL contribution to the muon anomaly,
$a_\mu=(g_\mu-2)/2$, is $(11\pm4)$ or $(10.5\pm2.6)\times 10^{-10}$
\cite{Bijnens:2007pz,Prades:2009tw,Jegerlehner:2009ry}
depending somewhat on which error estimates and which contributions are taken
into account. In this talk I have given an overview of a number of model
estimates with the emphasis on my old work 
\cite{Bijnens:1995cc,Bijnens:1995xf,Bijnens:2001cq} as well as a number of
newer developments. For the latter I have spent quite some time on our
reevaluation of the pion loop contribution
\cite{Bijnens:2016hgx,Bijnens:2012an,Abyaneh:2012ak,Amaryan:2013eja,Benayoun:2014tra}, as well as given a number of arguments why the HLS number
of \cite{Hayakawa:1995ps,Hayakawa:1996ki} should be considered obsolete.
The conclusion is that the pion-loop contributes
with $-(2.0\pm0.5)\times10^{-10}$.

One of the remaining problems in the model approach is that the class
of models with an ``unshielded'' quark-loop at relatively low-energies for the
photons tend to obtain larger numbers. Whether this is a real phenomenon or not
is a question which needs to be settled. My own opinion there is that I see no
counterpart of it in $\gamma\gamma\to$~hadrons at low to intermediate energies
beyond the already included single meson and two-pion exchanges.

For contributions of different mechanisms, progress can be expected both from
the dispersive approaches mentioned and experiment restricting the couplings
of off-shell or virtual photons to meson that go into the modeling.
Alternatively, a full new model calculation that includes phenomenology beyond
what the ENJL does, is very desirable as well as more work on the
short-distance aspects.
 
\section*{Acknowledgements}

I thank the organizers for the very nice atmosphere and the many
opportunities to discuss the issues regarding HLbL.
This work is supported in part by the Swedish Research Council grants
621-2013-4287, 2015-04089 and 2016-05996 and by
the European Research Council (ERC) under the European Union's Horizon 2020
research and innovation programme (grant agreement No 668679).

%
%

\end{document}